\documentclass[conference]{IEEEtran}
\usepackage{amsmath,amsfonts,amssymb,amsthm,thmtools}
\usepackage[ruled, linesnumbered, vlined]{algorithm2e}
\usepackage{algpseudocode}
\usepackage{array}
\usepackage[caption=false,font=normalsize,labelfont=sf,textfont=sf]{subfig}
\usepackage{textcomp}
\usepackage{stfloats}
\usepackage{url}
\usepackage{verbatim}
\usepackage{graphicx}
\usepackage{cite}
\usepackage{xcolor}
\usepackage{booktabs}
\usepackage{tabularx}
\usepackage{xspace}
\usepackage{tikz}
\usepackage{enumitem}
\usepackage{bm}
\usepackage{balance}
\usepackage{mathtools}
\usepackage{multirow}

\usetikzlibrary{automata, positioning, arrows, shapes, matrix, decorations.pathreplacing}

\newcommand\xqed[1]{%
  \leavevmode\unskip\penalty9999 \hbox{}\nobreak\hfill
  \quad\hbox{#1}}
\newcommand\extri{\xqed{$\triangle$}}

\algnewcommand\algorithmicforeach{\textbf{for each}}
\algdef{S}[FOR]{ForEach}[1]{\algorithmicforeach\ #1\ \algorithmicdo}

\SetKw{Continue}{continue}
\SetKw{Break}{break}

\def\sup#1{$^{#1}$}

\renewcommand{\baselinestretch}{0.99}
\begin{document}
\title{Practical Boolean Decomposition for Delay-driven LUT Mapping}


\author{
    \IEEEauthorblockN{
        Alessandro Tempia Calvino\sup{1}, Alan Mishchenko\sup{2},
        Giovanni De Micheli\sup{1}, Robert Brayton\sup{2}
    }

  \IEEEauthorblockA{
    \sup{1}Integrated Systems Laboratory, EPFL, Lausanne, Switzerland\\
    \sup{2}Department of EECS, University of California, Berkeley, USA
    }
}




\maketitle

\begin{abstract}
Ashenhurst-Curtis decomposition~(ACD) is a decomposition technique used, in particular, to map combinational logic into \emph{lookup tables}~(LUTs) structures when synthesizing hardware designs. However, available implementations of ACD suffer from excessive complexity, search-space restrictions, and slow run time, which limit their applicability and scalability. This paper presents a novel fast and versatile technique of ACD suitable for delay optimization. We use this new formulation to compute two-level decompositions into a variable number of LUTs and enhance delay-driven LUT mapping by performing ACD on the fly.
Experiments with heavily optimized benchmarks show an average delay improvement of $\mathbf{12.39}$\% and an area reduction of $\mathbf{2.20}$\% compared to state-of-the-art LUT mapping, with affordable run time. Additionally, our method improves the best-known delay for $\mathbf{4}$ benchmarks in the EPFL synthesis competition.

\end{abstract}


\IEEEpeerreviewmaketitle

\section{Introduction}
Ashenhurst-Curtis decomposition~(ACD)~\cite{Ashenhurst_1957, Curtis_1962}, also known as Roth-Karp decomposition~\cite{Roth_1962}, is a powerful technique that finds a decomposition of a Boolean function into a set of sub-functions and a composition function with reduced support. ACD finds applications in logic optimization and technology mapping. The noteworthy use cases of ACD are in mapping into standard cells~\cite{Kravets_2001} and \emph{field-programmable gate arrays}~(FPGA)~\cite{Legl_1998}, decomposition of multi-valued relations~\cite{Perkowski_1997}, and encoding of multi-valued networks~\cite{Jiang_2001}.

\begin{figure}[t]
    \centering
    \begin{tikzpicture}[scale=0.9, every node/.style={scale=0.9}]

    \node[anchor=base] (l) at (1, -0.6) {Bound set};
    \node[anchor=base] (l) at (3.6, -0.6) {Shared set};
    \node[anchor=base] (l) at (5.5, -0.6) {Free set};

     \node[rectangle, xshift=5.1cm, yshift=2.85cm, minimum height=1cm, minimum width=1.6cm, draw=black] (l1) {$L_1$};

    \node[rectangle, xshift=0cm, yshift=0.9cm, minimum height=1cm, minimum width=1.6cm, draw=black] (l2) {$L_2$};

    \node[rectangle, xshift=2cm, yshift=0.9cm, minimum height=1cm, minimum width=1.6cm, draw=black] (l3) {$L_3$};

    \draw (-0.6, 0) edge (-0.6, 0.4);
    \draw (-0.3, 0) edge (-0.3, 0.4);
    \draw (0.0, 0) edge (-0.0, 0.4);
    \draw (0.3, 0) edge (0.3, 0.4);
    \draw (0.6, 0) edge (0.6, 0.4);

    \node[anchor=base] (l) at (-0.6, -0.2) {\footnotesize $x_0$};
    \node[anchor=base] (l) at (-0.3, -0.2) {\footnotesize $x_1$};
    \node[anchor=base] (l) at (0.0, -0.2) {\footnotesize $x_2$};
    \node[anchor=base] (l) at (0.3, -0.2) {\footnotesize $x_3$};
    \node[anchor=base] (l) at (0.6, -0.2) {\footnotesize $x_4$};

    \draw (1.4, 0) edge (1.4, 0.4);
    \draw (1.7, 0) edge (1.7, 0.4);
    \draw (2.0, 0) edge (2.0, 0.4);
    \draw (2.3, 0) edge (2.3, 0.4);

    \node[anchor=base] (l) at (1.4, -0.2) {\footnotesize $x_0$};
    \node[anchor=base] (l) at (1.7, -0.2) {\footnotesize $x_1$};
    \node[anchor=base] (l) at (2.0, -0.2) {\footnotesize $x_2$};
    \node[anchor=base] (l) at (2.3, -0.2) {\footnotesize $x_3$};

    \draw (3.6, 0) edge (3.6, 1.4);
    \draw (2.6, 0.2) edge (3.6, 0.2);
    \draw (2.6, 0.2) edge (2.6, 0.4);
    \node[draw, shape=circle, fill=black, minimum size = 2pt, inner sep=0pt, xshift=3.6cm, yshift=0.2cm] (c0) {};
    \draw (3.6, 1.4) edge (5.1, 1.4);
    \draw (5.1, 1.4) edge (5.1, 2.35);

    \node[anchor=base] (l) at (3.6, -0.2) {\footnotesize $x_5$};

    \draw (5.4, 0) edge (5.4, 2.35);
    \draw (5.7, 0) edge (5.7, 2.35);

    \node[anchor=base] (l) at (5.4, -0.2) {\footnotesize $x_6$};
    \node[anchor=base] (l) at (5.7, -0.2) {\footnotesize $x_7$};

    \draw (0, 1.4) edge (0, 2);
    \draw (0, 2) edge (4.5, 2);
    \draw (4.5, 2) edge (4.5, 2.35);

    \draw (2, 1.4) edge (2, 1.7);
    \draw (2, 1.7) edge (4.8, 1.7);
    \draw (4.8, 1.7) edge (4.8, 2.35);

    \draw (l1) edge (5.1, 3.55);
\end{tikzpicture}
    \caption{ACD of an $8$-input Boolean function into three $5$-input LUTs with a $5$-variable \emph{bound set}~(BS), a $1$-variable \emph{shared set}~(SS), and a $2$-variable \emph{free set}~(FS).
    }
    \label{fig:acd}
\end{figure}
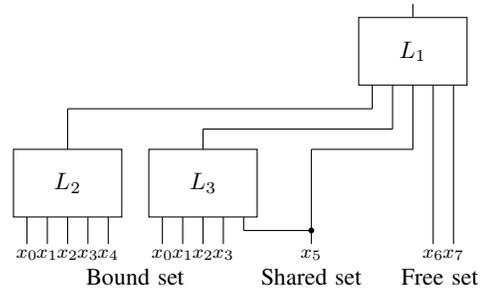

Traditional applications rely on the original formulation of ACD~\cite{Ashenhurst_1957, Curtis_1962, Roth_1962}, breaking the input variables into two groups: the bound set (BS) and the free set (FS). Other approaches to ACD~\cite{Legl_1998} allow for a shared set (SS) when one or more LUTs in terms of the BS variables are single-variable functions (buffers). The larger the SS size, the fewer LUTs are required.
For instance, Figure~\ref{fig:acd} shows an ACD of a function with BS, FS, and SS resulting in three $5$-input LUTs. In~\cite{Legl_1998}, maximizing the SS is implemented using \emph{binary decision diagrams}~(BDDs)~\cite{Bryant_86}. More recently, truth-table-based implementations eliminated the need for explicitly constructing a BDD, resulting in a faster decomposition~\cite{Mishchenko_2008, Ray_2012}.

ACD has been applied to map into fixed \emph{lookup table}~(LUT) structures~\cite{Ray_2012} as a way to mitigate structural bias and improve the quality of standard LUT mapping. This approach utilizes heuristic variable re-ordering to find an ACD, supporting up to $1$ SS variable. Additionally, ACD has been used in post-mapping resynthesis~\cite{Mishchenko_2008}, when logic cones composed of several LUTs are collapsed into single-output Boolean functions and re-expressed using fewer LUTs. The authors proposed to use \emph{disjoint-support} decomposition~(DSD) and Shannon's expansion to pack logic into LUTs while supporting up to $3$ SS variables.

Since ACD is often applied only to functions up to $11$ or $16$ inputs (for LUT structures composed of two or three $6$-LUTs, respectively), state-of-the-art LUT mapping is performed through local substitutions applied to an initial graph representation, called \emph{subject graph}. Generally, delay-optimal mapping w.r.t. the subject graph is feasible in polynomial time~\cite{Cong_1994}, while area-optimal mapping is NP-hard~\cite{329262}. However, the structure of the subject graph highly impacts the result. This phenomenon is known as \emph{structural bias}. To mitigate structural bias, methods in the literature generate a set of structural choices (or decompositions) available during mapping~\cite{644605, Chen_2001, 1560122}.


This paper offers two main contributions. First, we revisit the formulation of ACD with SS to enhance its computationally efficiency in LUT mappers and post-mapping resynthesis engines performing delay optimization. Based on the ideas presented in~\cite{Mishchenko_2023}, our algorithm is truth-table-based and flexible in the number of FS, BS, and SS variables, and in the number of BS functions. Our ACD runs up to $2$x faster, compared to~\cite{Ray_2012}, and up to $80$x faster, compared to~\cite{Mishchenko_2008} when performing decompositions into two $6$-LUTs. Furthermore, it also finds considerably more solutions.

Second, we use ACD for the delay optimization of LUT networks. The idea is to compute functional decompositions using the timing-critical variables in the FS and the rest of the variables in the BS and SS. We integrate our ACD into the state-of-the-art LUT mapper for delay optimization. To our knowledge, this is the first practical and scalable work that uses ACD for delay-driven LUT mapping.

We experimentally evaluate the performance of ACD and compare mapping based on Boolean decomposition against state-of-the-art methods:
\begin{enumerate}
    \item We compare our ACD method against other decomposition methods in ABC, showing better quality with a competitive or better run time.
    \item We demonstrate that mapping with ACD can efficiently mitigate structural bias and considerably reduce the delay. We compare the default LUT mapper in ABC, the LUT mapper with Boolean decomposition in ABC, and the proposed mapper with integrated ACD. We show that mapping with ACD outperforms the other mappers in delay by $7.52\%$ on average with and without structural choices~\cite{1560122}. Moreover, we show that an additional mapping round using the network obtained by ACD as a structural choice can further improve the delay, compared to the standard LUT mapper, by $12.39$\% with an area reduction of $2.20$\%.
    \item We present $4$ new best results in the EPFL competition.
\end{enumerate}

\section{Preliminaries} \label{sec:background}
This section introduces the basic notations and background related to logic networks, decomposition, and LUT mapping.

\subsection{Definitions}
A \textit{Boolean function} is a mapping from a $k$-dimensional 
Boolean space into a $1$-dimensional one: $\{0,1\}^k \rightarrow \{0,1\}$.

A \emph{truth table} representation of a $k$-input Boolean function $f:\{0,1\}^k\rightarrow\{0,1\}$ can be encoded as a bit string $b = b_{l-1} \dots b_0$, i.e., a sequence of bits, of length $l = 2^{k}$. A bit $b_i \in \{0, 1\}$ at position $0 \le i < l$ is equal to the value taken by $f$ under the input assignment $\vec{a} = (a_0, \dots, a_{k-1})$ where
\begin{equation*}
    2^{k-1} \cdot a_{k-1} + \dots + 2^0 \cdot a_0 = i.
\end{equation*}

The \emph{positive cofactor} of a Boolean function $f$ with respect to a variable $x_i$, represented as $f_{x_i}$, is the Boolean function obtained by setting $x_i = 1$. Similarly, the \emph{negative cofactor} $f_{\bar x_i}$  is the Boolean function obtained by setting $x_i = 0$.

In the classical representation, we refer to the leftmost input column of a truth table as the \emph{most significant variable} ($a_{k-1})$ and the rightmost input column as the \emph{least significant variable} ($a_{0})$. A \emph{swap} of two variables results in the interchange of the corresponding two-variable cofactors, thereby altering the truth table.

Figure~\ref{fig:tt} depicts two truth tables represented as bit strings, one in binary and one in hexadecimal. Notably, the rightmost truth table can be derived from the leftmost one by swapping the variables $x_0$ and $x_2$. Marked next to both truth tables are the cofactors with respect to two most significant variables.

\begin{figure}[t]
    \centering
    \begin{tikzpicture}[scale=0.9, every node/.style={scale=0.9}]
    \matrix (m1) [matrix of math nodes, inner sep=2pt, outer sep=0pt] at (0, 0) {
        x_2 & x_1 & x_0 & [0.5em] f \\[0.1em]
        0   &   0 &   0 &  1 \\
        0   &   0 &   1 &  0 \\
        0   &   1 &   0 &  1 \\
        0   &   1 &   1 &  0 \\
        1   &   0 &   0 &  1 \\
        1   &   0 &   1 &  1 \\
        1   &   1 &   0 &  0 \\
        1   &   1 &   1 &  1 \\
        };

    \draw ([xshift=-0.5em, yshift=0.1em]m1-1-4.north west) -- ([xshift=-0.5em, yshift=0.1em]m1-9-4.south west);
    \draw ([xshift=0, yshift=0.1em]m1-2-1.north west) -- ([xshift=0.5em, yshift=0.1em]m1-2-4.north east);

    \node[anchor=base] (l1) at (0, -2.1) {$f = 10110101$};

    \draw[decoration={brace,mirror,raise=5pt},decorate] (0.9,0.5) -- (0.9,1.2);
    \node[anchor=west] (b1) at (1.1, 0.85) {\footnotesize $f_{\bar x_1\bar x_2}$};

    \draw[decoration={brace,mirror,raise=5pt},decorate] (0.9,-0.25) -- (0.9,0.45);
    \node[anchor=west] (b2) at (1.1, 0.1) {\footnotesize $f_{x_1 \bar x_2}$};

    \draw[decoration={brace,mirror,raise=5pt},decorate] (0.9,-1) -- (0.9,-0.3);
    \node[anchor=west] (b3) at (1.1, -0.65) {\footnotesize $f_{\bar x_1  x_2}$};

    \draw[decoration={brace,mirror,raise=5pt},decorate] (0.9,-1.75) -- (0.9,-1.05);
    \node[anchor=west] (b3) at (1.1, -1.4) {\footnotesize $f_{x_1 x_2}$};

    \node[anchor=base] (s) at (2.2, 1.6) {\footnotesize $x_0 \leftrightarrow x_2$};
    \draw [->, line width=0.3mm] (1.4, 1.4) -- (3, 1.4);

    \matrix (m2) [matrix of math nodes, inner sep=2pt, outer sep=0pt] at (4.4, 0) {
        x_0 & x_1 & x_2 & [0.5em] f \\[0.1em]
        0   &   0 &   0 &  1 \\
        0   &   0 &   1 &  1 \\
        0   &   1 &   0 &  1 \\
        0   &   1 &   1 &  0 \\
        1   &   0 &   0 &  0 \\
        1   &   0 &   1 &  1 \\
        1   &   1 &   0 &  0 \\
        1   &   1 &   1 &  1 \\
        };

    \draw ([xshift=-0.5em, yshift=0.1em]m2-1-4.north west) -- ([xshift=-0.5em, yshift=0.1em]m2-9-4.south west);
    \draw ([xshift=0, yshift=0.1em]m2-2-1.north west) -- ([xshift=0.5em, yshift=0.1em]m2-2-4.north east);

    \node[anchor=base] (l2) at (4.4, -2.1) {$f =$ 0\text{x}A$7$};

    \draw[decoration={brace,mirror,raise=5pt},decorate] (5.3,0.5) -- (5.3,1.2);
    \node[anchor=west] (b1) at (5.5, 0.85) {\footnotesize $f_{\bar x_0\bar x_1}$};

    \draw[decoration={brace,mirror,raise=5pt},decorate] (5.3,-0.25) -- (5.3,0.45);
    \node[anchor=west] (b2) at (5.5, 0.1) {\footnotesize $f_{\bar x_0 x_1}$};

    \draw[decoration={brace,mirror,raise=5pt},decorate] (5.3,-1) -- (5.3,-0.3);
    \node[anchor=west] (b3) at (5.5, -0.65) {\footnotesize $f_{x_0 \bar x_1}$};

    \draw[decoration={brace,mirror,raise=5pt},decorate] (5.3,-1.75) -- (5.3,-1.05);
    \node[anchor=west] (b3) at (5.5, -1.4) {\footnotesize $f_{x_0 x_1}$};

\end{tikzpicture}
    \caption{Truth table representations and their encoding, cofactor extraction w.r.t. the two most significant variables, and variable swapping of $x_0$ with~$x_2$.}
    \label{fig:tt}
\end{figure}

A completely specified Boolean function $f$ \emph{essentially depends} on a variable $v$ if there exists an input combination such that the value of the function changes when the variable is toggled ($\frac{\partial f}{\partial v}=1$). The \emph{support of} $f$ is the set of all variables on which function $f$ essentially depends. The supports of two functions are \emph{disjoint} if they do not contain common variables. A set of functions is disjoint if their supports are pair-wise disjoint.


A \emph{Boolean network} is modeled as a directed acyclic graph~(DAG) with nodes represented by Boolean functions. The sources of the graph are the \emph{primary inputs}~(PIs), the sinks are the \emph{primary outputs}~(POs). For any node $n$, the \textit{fanins} of $n$ is a set of nodes driving $n$, i.e. nodes that have an outgoing edge towards $n$. Similarly, the \textit{fanouts} of $n$ is a set of nodes driven by node $n$, i.e., nodes that have an incoming edge from $n$.
A $k$\emph{-LUT network} is a Boolean network composed of $k$-input \emph{lookup tables}~($k$-LUTs) capable of realizing any $k$-input Boolean function.
An \emph{and-inverter graph}~(AIG)~\cite{Mishchenko_Brayton_2006} is a Boolean network where nodes are $2$-input ANDs and edges may implement inverters.

A \emph{cut} $C$ of a Boolean network is a pair ($n$, $\mathcal{K}$), where $n$ is a node called \emph{root}, and $\mathcal{K}$ is a set of nodes, called \emph{leaves}, such that 1) every path from any PI to node $n$ passes through at least one leaf and 2) for each leaf $v \in \mathcal{K}$, there is at least one path from a PI to $n$ passing through $v$ and not through another leaf. The \emph{size} of a cut is the number of leaves. A cut is $k$-feasible if its size does not exceed $k$.

\subsection{Ashenhurst-Curtis decomposition}
Ashenhurst-Curtis decomposition~(ACD)~\cite{Ashenhurst_1957, Curtis_1962, Roth_1962}, of a single-output Boolean function $f$ can be expressed as follows:
\begin{equation} \label{eq:acd}
    f(\vec x_{bs}, \vec x_{ss}, \vec x_{fs}) = g( \vec h(\vec x_{bs}, \vec x_{ss}), \vec x_{ss}, \vec x_{fs} ),
\end{equation}
where $\vec x_{bs}$ is the \emph{bound set}~(BS), $\vec x_{ss}$ is \emph{shared set}~(SS), and $\vec x_{fs}$ is the \emph{free set}~(FS). These sets are disjoint variable subsets, which together form the support of $f$. The function $\vec h$ may be multi-output with the number of outputs less than the BS size. The single-output functions in $\vec h$ are referred to as BS functions. The function $g$ is referred to as the \emph{composition function}. When decomposing into $k$-LUTs, the composition function is typically chosen to fit into one $k$-input LUT. Figure~\ref{fig:acd} shows an ACD of an $8$-input function into three $5$-input LUTs with a $5$-variable BS, a $1$-variable SS, and a $2$-variable FS. The decomposition generates two BS functions~($L_2$, $L_3$) and a composition function~($L_1$) with $5$ inputs.

\subsection{FPGA technology mapping}
LUT mapping is the process of expressing a Boolean network in terms of $k$-input lookup tables ($k$-LUTs).
Before mapping, the network is represented as a \textit{k-bounded} Boolean network called the \textit{subject graph}, which contains nodes with a maximum fanin size of \textit{k}. The AIG is the most common subject graph representation. The subject graph is transformed into a mapped network by applying local substitutions to sections of the circuit defined by cuts, which are computed using cut enumeration~\cite{10.1145/296399.296425}. A LUT mapper computes a mapping solution by selecting a subset of the cuts that cover the subject graph while minimizing a cost function. The state-of-the-art LUT mapper computes cuts and refines the mapping solution in several mapping passes using heuristics based on delay, area, and edge count. For further details, refer to~\cite{4397290}.

\section{Improvements to ACD} \label{sec:acd}

\begin{figure}[t]
    \centering
    \begin{tikzpicture}[scale=0.9, every node/.style={scale=0.9}]
    \node[anchor=base] (a) at (2.3, 3.6) {\footnotesize{$00$}};
    \node[anchor=base] (b) at (3.5, 3.6) {\footnotesize{$01$}};
    \node[anchor=base] (c) at (4.7, 3.6) {\footnotesize{$10$}};
    \node[anchor=base] (d) at (5.9, 3.6) {\footnotesize{$11$}};

    \node[anchor=base] (f) at (3.7, 5.2) {$f$};
    \node[anchor=base] (l) at (6.4, 4.3) {$L_1$};

    \node[anchor=base] (l) at (1, -0.5) {Bound set};
    \node[anchor=base] (l) at (3.2, -0.5) {Shared set};
    \node[anchor=base] (l) at (5.3, -0.5) {Free set};



    \node[trapezium, xshift=4.1cm, yshift=4cm, minimum height=1cm, minimum width=5cm, draw=black, trapezium stretches, trapezium left angle=80, trapezium right angle=80] (c1) {};

    \node[rectangle, xshift=3.7cm, yshift=3.35cm, minimum height=2.7cm, minimum width=6cm, draw=black, dashed] (l1) {};

    \node[rectangle, xshift=1cm, yshift=1cm, minimum height=1.2cm, minimum width=2cm, draw=black] (l2) {$L_2$};

    \node[rectangle, xshift=2.3cm, yshift=2.8cm, minimum height=0.6cm, minimum width=1cm, draw=black] (g0) {$g_{0}$};

    \node[rectangle, xshift=3.5cm, yshift=2.8cm, minimum height=0.6cm, minimum width=1cm, draw=black] (g1) {$g_{1}$};

    \node[rectangle, xshift=4.7cm, yshift=2.8cm, minimum height=0.6cm, minimum width=1cm, draw=black] (g2) {$g_{2}$};

     \node[rectangle, xshift=5.9cm, yshift=2.8cm, minimum height=0.6cm, minimum width=1cm, draw=black] (g3) {$g_{3}$};



    


    \draw (0.4, 0.4) edge (0.4,0);
    \draw (0.8, 0.4) edge (0.8,0);
    \draw (1.2, 0.4) edge (1.2,0);
    \draw (1.6, 0.4) edge (1.6,0.2);

    \draw (1, 1.6) edge (1, 4.2);
    \draw (1, 4.2) edge (2.06, 4.2);

    \draw (3, 0) edge (3, 1.8);
    \draw (1.6, 0.2) edge (3, 0.2);
    \node[draw, shape=circle, fill=black, minimum size = 2pt, inner sep=0pt, xshift=3cm, yshift=0.2cm] (c3) {};
    \draw (3, 1.8) edge (1.4, 1.8);
    \draw (1.4, 1.8) edge (1.4, 3.9);
    \draw (1.4, 3.9) edge (1.86, 3.9);

    \draw (g0) edge (2.3, 3.5);
    \draw (g1) edge (3.5, 3.5);
    \draw (g2) edge (4.7, 3.5);
    \draw (g3) edge (5.9, 3.5);

    \draw (4.5, 0) edge (4.5, 2.5);
    \node[draw, shape=circle, fill=black, minimum size = 2pt, inner sep=0pt, xshift=4.5cm, yshift=2.2cm] (c0) {};
    \draw (5.7, 2.2) edge (2.1, 2.2);
    \draw (5.7, 2.2) edge (5.7, 2.5);
    \node[draw, shape=circle, fill=black, minimum size = 2pt, inner sep=0pt, xshift=3.3cm, yshift=2.2cm] (c0) {};
    \draw (2.1, 2.2) edge (2.1, 2.5);
    \draw (3.3, 2.2) edge (3.3, 2.5);

    \draw (6.1, 0) edge (6.1, 2.5);
    \node[draw, shape=circle, fill=black, minimum size = 2pt, inner sep=0pt, xshift=6.1cm, yshift=2.35cm] (c0) {};
    \draw (6.1, 2.35) edge (2.5, 2.35);
    \draw (2.5, 2.35) edge (2.5, 2.5);
    \node[draw, shape=circle, fill=black, minimum size = 2pt, inner sep=0pt, xshift=3.7cm, yshift=2.35cm] (c0) {};
    \draw (3.7, 2.35) edge (3.7, 2.5);
    \draw (4.9, 2.35) edge (4.9, 2.5);
    \node[draw, shape=circle, fill=black, minimum size = 2pt, inner sep=0pt, xshift=4.9cm, yshift=2.35cm] (c0) {};

    \draw (l1) edge (f);

\end{tikzpicture}
    \caption{Illustrating the AC decomposition of a Boolean function}
    \label{fig:acd_mux}
\end{figure}
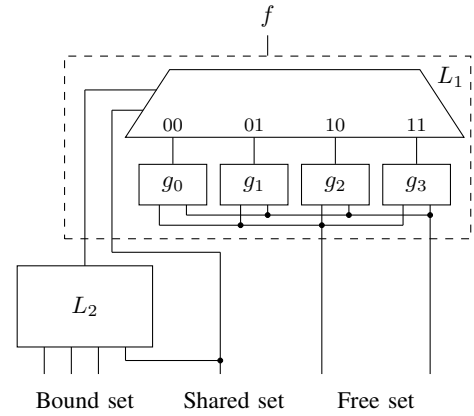

This section discusses a fast and versatile truth-table-based implementation of ACD for single-output functions with support for a shared set. We propose several novelties that make ACD practical within LUT mappers and resynthesis methods.
Figure~\ref{fig:acd_mux} illustrates the ACD computation. The BS, SS, FS, and the number of BS functions used are flexible and determined during the decomposition. The composition function~($L_1$) is implemented as a multiplexer of cofactors with respect to BS functions and the shared set. Functions dependent on the FS~($g_{i})$, called FS functions, are the data inputs of the multiplexer found inside the composition function. BS functions and the shared set are instead the selection inputs.

This definition of decomposition reflects the one used by previous approaches~\cite{Legl_1998}. Specifically, the decomposition is generic, i.e., it includes other types of decomposition. For instance, a Shannon's expansion:
\begin{equation*}
    f = x f_{x} + \bar x f_{\bar x},
\end{equation*}
where $x$ is a selector of a multiplexer, can be re-expressed in our ACD format:
\begin{equation*}
    f = f_{x} f_{\bar x} 1  + f_{x} \bar f_{\bar x} x + \bar f_{x} f_{\bar x} \bar{x} + \bar f_{x} \bar f_{\bar x} 0,
\end{equation*}
where $x$ is a FS variable, $f_{x}$ and $f_{\bar x}$ are BS functions, and FS fuctions $g_i$ are $1$, $x$, $\bar x$, and $0$.


In this section, we first present how to efficiently check the existence of a feasible ACD and assign variables to the FS, BS, and SS (Section~\ref{sec:FSBS}). Next, we show how to compute the decomposition while minimizing the number of BS functions and their support (Section~\ref{sec:encoding}).


\subsection{Finding a feasible decomposition} \label{sec:FSBS}
After defining the properties of ACD, in this section we present an efficient method to check the existence of a Boolean decomposition and find an assignment of support variables to the FS and the BS (and SS). In particular, we focus on decomposition into two levels of $k$-input LUTs. For simplicity, in this section we consider SS variables a part of the BS.

The first step to derive a decomposition is to partition of variables into FS and BS. Given a truth table, our approach enumerates different free sets. Let $N$ be the number of variables in the support of a function to decompose. Let $P$ be the number of variables to consider in the FS. The remaining $N-P$ variables are considered in the BS. The number of different free sets is $\binom{N}{P}$. Regarding the choice of value $P$ when searching for a feasible two-level decomposition, for an $N$-input function and $k$-input LUTs, it is convenient to consider ($N-k$) variables in the FS, so that at most $k$ variables are considered in the BS. For instance, when $N=8$ and $k=6$, we can choose $P=2$ and evaluate $8 \cdot 7/2 = 28$ different $2$-variable free sets.

For each FS, the truth table is transformed to have the FS variables as the least significant ones, compared to the BS variables. The variable reordering is performed using a dedicated procedure, which swaps two variables. Note that when enumerating all the free sets the first FS composed of the P least significant variables in the support of the function does not need variable swapping since the original truth table already reflects this order. Then, every consecutive FS can be derived from a previous FS by swapping one variable in $x_{fs}$ with one in $x_{bs}$. The complexity to explore all the FS is of $\binom{N}{P}$ swap operations. Figure~\ref{fig:tt} shows how a variable swap affects the truth table.

Each input assignment to the BS variables selects one $P$-input function in terms of the FS variables. Specifically, each $P$-input function is a cofactor with respect to $x_{bs}$. From a truth table in this format, FS functions are easily computed by extracting groups of $2^{P}$ bits at $i\cdot 2^{P}$ offsets with $i\in[0,2^{(N-P)})$. Informally, FS functions are listed next to each other. Figure~\ref{fig:tt} graphically depicts the extraction of cofactors with respect to the two most significant variables.

\emph{Example 1:} Let us consider the $6$-variable function represented in hexadecimal format as a truth table $f =$ 0x8804800184148111. Let us assume that the FS variables are the two least significant variables and the BS variables are the four most significant ones. The functions in terms of FS variables have truth tables with $2^P = 2 ^ 2 = 4$ bits. There are $2^{(N-P)} = 16$ of them, corresponding to hexadecimal digits in the truth table (0x8, 0x8, 0x0, 0x4, etc). \extri

The target function can be realized using $M$ bound set functions if the number of unique FS functions, known as column multiplicity $\mu$, does not exceed $2^M$, hence $M \ge \lceil \log_2(\mu) \rceil$. If $P + M \le k$, the composition function can be implemented as a $k$-LUT.

\emph{Example 2:} Continuing Example~1, there are $16$ FS functions of which only $4$ are unique. The FS functions are 0x8, 0x0, 0x4, and 0x1. Hence, the column multiplicity $\mu = 4$, which needs at least $M = \lceil \log_2(4) \rceil = 2$ BS functions. Hence, this partition of variables into FS and BS produces a feasible support-reducing decomposition into $4$-input LUTs. Using Figure~\ref{fig:acd_mux}, ACD assigns FS functions to $g_{i}$. Then, two BS functions of at most $4$ inputs are necessary to select the correct FS function. \extri

We employ the enumeration of free sets while counting the number of unique cofactors to check if a support-reducing decomposition exists. In practice, a sufficient condition for a $2$-level decomposition is to have $M+P \le k$ and $N-P \le k$, i.e., the composition function is $k$-feasible, and the number of remaining variables in the BS does not exceed $k$.


After identifying a partition of variables into FS and BS, and the corresponding unique FS functions, our method uses the techniques in Section~\ref{sec:encoding} to produce a decomposition while minimizing the number of BS functions and their support.

\subsection{Functional encoding and support minimization} \label{sec:encoding}
Once a partition of variables into FS and BS with a feasible decomposition is found, the BS functions are extracted by assigning each FS function to an encoding. Informally, an encoding represents the assignment of FS functions to the data inputs of the MUX of Figure~\ref{fig:acd_mux} (e.g., the encoding of $g_1$ is $01$). While any encoding that distinguishes FS functions is a valid solution, a good encoding also minimizes the number of BS functions required (by maximizes the shared set), and the functional support. In particular, it is crucial to find an encoding that minimizes the support for three reasons. First, if $N-P > k$, by minimizing the support, each BS function would ideally fit into a $k$-LUT, and the decomposition is feasible in $2$ levels. Second, minimizing the support maximizes the shared set (buffer BS functions), reducing the number of required LUTs. Third, the number of edges required is reduced, helping routability.
Finding a feasible encoding is similar to solving constrained encoding problems~\cite{DeMicheli_1985, Villa_1990, Yang_1991}.

An encoding is an assignment of a code $T = t_{M-1}\dots t_0$ of length $M$ to each FS function. A variable $t_i$ takes one of the three values, $1$, $0$, or $-$, indicating the ON-set, OFF-set, and DC-set, respectively.
Let \emph{i-sets} be the set of $\mu$ Boolean functions in terms of the BS variables encoding FS functions using one-hot encoding. Precisely, an i-set represents one FS function and takes value $1$ when an input assignment to the BS variables results in the corresponding FS function.

\emph{Example 3}: Using Example 2, the i-set corresponding to the FS function 0x8 is 1100100010001000 in binary format. Note that the truth table has $N-P$ variables and has value $1$ when the original function is 0x8. \extri

I-sets are used to derive a more compact encoding with a two-step procedure. The first one enumerates \emph{candidate BS functions}. The second one solves a unate covering problem in which columns are candidate BS functions and rows are pairs of FS functions to be distinguished.

Candidate BS functions are functions depending on BS variables whose output can used as $t_i$ to encode FS functions. They are enumerated by combining i-sets. To leverage all the functional degrees of freedom of a strict encoding, i-sets in a BS candidate can be either in the \emph{ON-set}, \emph{OFF-set}, or \emph{don't-care}~(DC) set. Since candidate BSs are used as select inputs of a multiplexer, BS candidates can distinguish elements in the ON-set (takes value $1$) against elements in the OFF-set (takes value $0$). In encoding problems, BS functions are called \emph{dichotomies}, while the pairs of functions to be distinguished are referred to as \emph{seed dichotomies}~\cite{Yang_1991}. Don't-cares in BS candidates are also important to minimize the support, which translates into fewer LUT edges.

\emph{Example 4}: Continuing Example 3, let us consider the candidate bound set function $h$ that has the i-sets \{0x8, 0x1\} in the ON-set and the i-set \{0x4\} in the OFF-set. Its function in binary format is $h=$11-01{-}{-}110101111 where ``-'' is a don't care. When $h=1$, either 0x8 or 0x1 are selected. When $h=0$, 0x4 is selected. The corresponding dichotomy is \{\{0x8, 0x1\},\{0x4\}\}. In this case, function $h$ distinguishes 0x8 from 0x4 and 0x1 from 0x4, covering the two seed dichotomies \{\{0x8\},\{0x4\}\} (or \{\{0x4\},\{0x8\}\}) and \{\{0x1\},\{0x4\}\} (or \{\{0x4\},\{0x1\}\}). \extri

A candidate bound set function is generated by assigning each i-set to be in the ON-set, OFF-set, or DC-set. Hence, the total number of possible BS candidates is $3^{\mu}$. Nonetheless, some BS candidates are interchangeable, i.e., one candidate can be obtained by swapping the ON-set and the OFF-set of another BS candidate. Our enumeration removes these symmetries by fixing one i-set to be only in the ON-set or DC-set, enumerating only $2\cdot 3^{\mu-1}$ BS candidates. Moreover, candidates not distinguishing any pair of FS functions are removed. As a special case, if $\mu$ is a power of $2$, the number of possible BS candidates reduces to $\binom{M}{M/2}/2$ by splitting the FS functions to be equally distributed between ON-set and OFF-set, i.e., each BS candidate must distinguish half of the FS functions against the other half. 

One limitation of this method is that the number of BS candidates is exponentially dependent on the column multiplicity. However, we may further reduce the number of BS candidates when it is too large. In particular, for an ACD into $6$-LUTs the maximum column multiplicity to support is $16$.
Consequently, the highest number of BS candidates is $9.5$ million for $\mu=15$. To maintain a reasonable number of BS candidates, our method does not use don't cares for problems with $\mu > 8$, enumerating $2^{\mu - 1}$ candidates and reducing the highest number of candidates to $16$ thousand. Through experimentation, we have observed that imposing this limitation scarcely affects the quality of the encoding, while substantially enhancing run-time efficiency. Conversely, extending this method to lower multiplicity values noticeably compromises the solution quality.

Each BS candidate function is associated with a cost that depends on the number of variables in its support. The number of variables is computed with a special procedure that considers don't cares.
Then, a covering table is constructed by having all the pairs of FS functions to be distinguished (seed dichotomies) as rows and the BS candidates as columns. A row-column entry $(i,j)$ is $1$ if the BS candidate of column $j$ distinguishes the seed dichotomy $i$. A solution that minimizes the support is computed by solving a minimum-cost covering problem~\cite{Yang_1991}. The solution must cover all the rows while minimizing the cost. We use greedy covering followed by local search to compute cost-minimizing cover. A single iteration of greedy covering extracts one column covering the most non-covered rows while minimizing the cost. The process is iterated until a solution is found. Then, the solution is iteratively improved by replacing one column with another having a lower cost.

\emph{Example 5:} Figure~\ref{fig:cover} shows a covering table reflecting the examples in this section. Each column in the table is a candidate BS function shown as a truth table in hexadecimal format on $4$ variables. Each BS candidate has a cost based on the number of variables on its support. Each row is a seed dichotomy. An element $(i,j)$ in the table is $1$ if the BS$_j$ distinguishes the seed dichotomy $i$. The best solution with cost $6$ takes the second and third columns and results in two BS functions depending on $3$ variables. \extri

\begin{figure}[t]
    \centering
    \begin{tikzpicture}[scale=0.9, every node/.style={scale=0.9}]



    \matrix (m) [matrix of math nodes, inner sep=2pt, outer sep=0pt]{
        {}                   & [1em] 4 & 3 & 3 \\
        {}                   & \text{C}9\text{AF} & 1177 & 2727 \\[0.1em]
        \{\{0\text{x}8\}, \{0\text{x}0\}\} & 1 & 0 & 1 \\
        \{\{0\text{x}8\}, \{0\text{x}4\}\} & 1 & 1 & 0 \\
        \{\{0\text{x}8\}, \{0\text{x}1\}\} & 0 & 1 & 1 \\
        \{\{0\text{x}0\}, \{0\text{x}4\}\} & 0 & 1 & 1 \\
        \{\{0\text{x}0\}, \{0\text{x}1\}\} & 1 & 1 & 0 \\
        \{\{0\text{x}4\}, \{0\text{x}1\}\} & 1 & 0 & 1 \\
        };

    \draw ([xshift=0.5em, yshift=0.1em]m-3-1.north east) -- ([xshift=0.5em, yshift=0.1em]m-8-1.south east);
    \draw ([xshift=0.5em, yshift=0.1em]m-3-1.north east) -- ([xshift=0.5em, yshift=0.1em]m-3-4.north east);
\end{tikzpicture}
    \caption{Covering table to solve the encoding problem.}
    \label{fig:cover}
\end{figure}

Given a solution, an encoding of the FS functions is obtained by assigning a code $T = t_{M-1}\dots t_0$, in which each variable $t_i$ corresponds to a selected BS$_i$ candidate.

\emph{Example 6:} Continuing Example 5, a minimum cover involves BS$_0 =$ 0x1177, by taking 0x4 and 0x1 in the ON-set, and BS$_1 =$ 0x2727 by taking 0x0 and 0x1 in the ON-set. Given the BS functions, the encoding of the FS functions assigns the following codes to $g_{i}$ in Figure~\ref{fig:acd_mux}: $T_{0\text{x}8} =$ 00, $T_{0\text{x}4} =$ 01, $T_{0\text{x}0} =$ 10, and $T_{0\text{x}1} =$ 11. Finally, the composition function is computed using the FS and its encoding, resulting in function 0x1048 when represented in hexadecimal format. Consequently, the function has been successfully decomposed using three $4$-LUTs. \extri 





\section{Technology mapping with ACD} \label{sec:tmap}
In this section, we leverage the Ashenhurst-Curtis decomposition~(ACD) methods described in Section~\ref{sec:acd} to improve the delay of LUT networks. ACD can be used in two ways: 1) as part of LUT mapping or 2) as a post-mapping resynthesis method to compact logic and decrease the delay. In this work, we focus on the former usage since it has more flexibility and optimization opportunities. Although post-mapping resynthesis is not covered in this work, its implementation would follow a methodology similar to~\cite{Mishchenko_2008}.
First, this section discusses how to perform delay-oriented functional decomposition for any number of FS variables and BS functions. Then, it describes the integration of ACD in a technology mapper.


\subsection{Delay-oriented ACD} \label{sec:dalayACD}
Let us consider a node $n$ in a $k$-LUT network and a cut $C$ rooted in $n$ that contains leaves in the input sub-network of $n$. Among all the leaves, some are timing-critical and some are not. Let $D$ be the latest arrival delay of a leaf in $C$. We use ACD to find an implementation that realizes the function of cut $C$ with delay $D+1$ where $|C|>k$, assuming a unit-delay model. Specifically, we use the timing-critical leaves of $C$ in the FS and other non-critical ones in the BS or SS. This transformation may reduce the worst delay of a LUT network when applied on the critical path.


The ACD-based transformation is performed in two steps. First, our method verifies the existence of a delay-minimizing decomposition. Second, if a decomposition exists, it solves the encoding problem and returns a solution.

\subsubsection{Checking the existence of a decomposition}
Algorithm~\ref{alg:acd_eval} shows the procedure \emph{evaluate} to check the existence of an ACD. The algorithm receives the function represented as a truth table $tt$ of a large cut with size $N$ where $N>k$. Set $S$ contains a list of timing-critical variables with delay $D$. First, the truth table is transformed to have critical variables as the least significant ones since they must be in the FS (at line \ref{alg:vars_reorder}). The proposed approach limits $N-P \le k$ to ensure a two-level decomposition without solving the encoding problem. Hence, the number of variables in the FS must be at least $P \ge N - k$, and $P \ge |S|$ to include all the delay-critical variables (at line \ref{alg:fs_size}). For each FS of $P_i$ variables, the column multiplicity value is computed using the method described in Section~\ref{sec:FSBS}, and the smallest one is returned (at line \ref{alg:multiplicity}). In this case, since delay-critical variables are always part of the FS, $\binom{N}{P_i-|S|}$ different combinations are enumerated. If the smallest multiplicity found can be implemented using at most $k-P_i$ BS functions, a delay-minimizing ACD exists. In this case, variables in the FS have the delay increase of $1$ while other variables have the delay increase of $2$ (at line~\ref{alg:delay_profile}). If, on the other hand, a decomposition with $P_i$ does not exist, the function is not decomposable.

The loop in line~\ref{alg:fs_size} begins checking the existence of a decomposition with a smaller value of $P$. This approach is based on the theoretical property that if a function is not decomposable for the given value of $P$, it is also not decomposable for $P+1$.
Then, if a decomposition exists, the loop attempts to increase the number of variables in the free set. Specifically, maximizing the free set to include non-critical variables has multiple benefits. Primarily, the decomposition would have a reduced column multiplicity, which simplifies the encoding problem. Additionally, maximizing the free set may increase the required time of the associated non-critical signals, facilitating the area-recovery process of technology mapping.

\begin{algorithm}[t]
    \small
    \textbf{Input} $\;$: Truth table $tt$, LUT size $k$, Late vars set $S$\\
    \textbf{Output}: Propagation delay\\

    reorder\_variables($tt$, $S$)\; \label{alg:vars_reorder}
    $\mu_{best} \gets \infty$\;
    $x_{fs} \gets \emptyset$\;
    \For{$P_i \gets \max($num\_vars$(tt) - k$, $|S|)$ to $k - 1$}{ \label{alg:fs_size}
        $\{\mu, x_{fs}'\} \gets$ compute\_smallest\_multiplicity($tt$, $P_i$, $|S|$)\; \label{alg:multiplicity}
        \If {$\mu \le 2^{k - P_i}$ and $\mu < \mu_{best}$} {
            $\mu_{best} \gets \mu$\;
            $x_{fs} \gets x_{fs}'$\;
            \Continue\; \label{alg:continue}
        }
        \Break\;
    }

    \If {$\mu_{best} \neq \infty$ } {
        \Return compute\_propagation\_delay($tt$, $x_{fs}$)\; \label{alg:delay_profile}
    }
    \Return infinite\_propagation\_delay()\;
    \caption{ACD evaluation}\label{alg:acd_eval}
\end{algorithm}

\subsubsection{Computing the decomposition}
After applying \emph{evaluate}, another procedure \emph{decompose} is used to compute the actual decomposition using the methods described in Section~\ref{sec:encoding}. 

\subsection{LUT mapping with ACD} \label{sec:acd_mapping}
The methods described in Section~\ref{sec:dalayACD} have been integrated into the LUT mapping algorithm in~\cite{4397290}. Each mapping iteration computes $k$-feasible cuts rooted in nodes of the subject graphs and selects one best cut for each node based on cost functions and slack. 
Typically, enumerated cuts are $k$-feasible, i.e., any cut abstracts a $k$-LUT. In our implementation, cut enumeration computes large cuts up to size $k < l \le 11$, where $l$ is provided by the user. During cut enumeration, the mapper computes cut functions as truth tables. For the non-$k$-feasible computed cuts, the mapper uses Algorithm~\ref{alg:acd_eval} to check the existence of a delay-minimizing decomposition into $k$-LUTs. If a decomposition is not feasible, the cut is discarded. If a decomposition exists, the cut delay is computed using the propagation delay returned by Algorithm~\ref{alg:acd_eval}. The area is computed pessimistically, neglecting the existence of a shared set, i.e., $Area = \lceil \log_2{\mu} \rceil + 1$. To have precise area information, i.e., the number of required LUTs, ACD would need to solve the encoding problem and compute the decomposition. However, experimentally, not running the decomposition on the fly reduces the run time considerably with negligible impact on the final circuit area.

The mapper uses $l$-feasible cuts with ACD in the delay mapping pass, while it uses $k$-feasible cuts in the following area recovery iterations. Note that area-recovery aims at improving the solution over non-critical paths and can always re-use the best cuts from the previous pass, such that the required times are met.
After the last mapping pass, a cover is generated consisting of $k$- and $l$-feasible cuts. At this stage, the mapper decomposes the non-$k$-feasible cuts into $k$-LUTs.

\section{Experiments} \label{sec:experiments}
This section presents an experimental evaluation of the proposed LUT mapping with ACD. First, the ACD algorithm proposed in this paper is compared with other state-of-the-art methods for decomposing practical functions. Then, we evaluate ACD for delay-driven LUT mapping. While the experiments are reported for $6$-input LUTs, similar improvements have been obtained for $4$-input LUTs as well.

The proposed methods have been implemented in \emph{ABC}~\cite{ABC}. For our experiments, we use the EPFL combinational benchmark suite~\cite{Amar2015TheEC} containing several circuits provided as \emph{and-inverter graphs}~(AIGs).
The baseline has been obtained using the commands and scripts ``\texttt{dfraig; resyn; resyn2; resyn2rs; if -y -K 6; resyn2rs}'' in ABC, which perform a high-effort size and depth AIG optimization. In particular, it combines SAT sweeping~\cite{Mishchenko05fraigs:a}, scripts for delay-oriented AIG optimization~\cite{Mishchenko_Brayton_2006}, and lazy man's logic synthesis~\cite{Yang_2012}, which is the most aggressive depth minimization command in ABC. The experiments have been conducted on an Intel i$5$ quad-core $2$GHz on MacOS. The results have been verified using combinational equivalent checkering in ABC. We extended the LUT mapper \emph{if} in ABC to perform ACD as discussed in Section~\ref{sec:tmap}.
The following commands are used in the experiments:
\begin{itemize}
    \item \texttt{dch (-f)}: computes structural choices used to mitigate the structural bias~\cite{1560122}, where \texttt{-f} stands for ``fast'';
    \item \texttt{if -K 6}: performs delay-oriented technology mapping with choices into $6$-LUTs using $6$-feasible cuts;
    \item \texttt{if -s -S 66 -K 8}: performs delay-oriented technology mapping using $8$-feasible cuts and decomposes logic for minimal delay into two $6$-LUTs using a SAT-based formulation (available in ABC but not published);
    \item \texttt{if -Z 6 -K 8}: performs technology mapping into $6$-LUTs using the proposed implementation of  delay-oriented ACD described in Section~\ref{sec:tmap} for $8$-feasible cuts;
    \item \texttt{st}: derives an AIG from an LUT network.
\end{itemize}


\subsection{Decomposition success rate}
\begin{table*}[th]
\center
\setlength\tabcolsep{2pt}
\caption{Decomposition success ratio into two $6$-LUTs for practical functions using different ACD methods.}
\label{tab:acd_success}
\begin{tabularx}{\textwidth}{lXrrXrrXrrXrrXrr}
\toprule
\textbf{ACD type} &  & \multicolumn{2}{c}{\textbf{7 vars} (41071)} & \multicolumn{1}{l}{} & \multicolumn{2}{c}{\textbf{8 vars} (107466)} & \multicolumn{1}{l}{} & \multicolumn{2}{c}{\textbf{9 vars} (195602)} & \multicolumn{1}{l}{} & \multicolumn{2}{c}{\textbf{10 vars} (313649)} & \multicolumn{1}{l}{} & \multicolumn{2}{c}{\textbf{11 vars} (404991)} \\
         &  & Success (\%)      & Time(s)     &                      & Success (\%)      & Time(s)     &                      & Success (\%)      & Time(s)     &                      & Success (\%)      & Time(s)      &                      & Success (\%)      & Time(s)      \\ \cline{1-1} \cmidrule{3-4} \cmidrule{6-7} \cmidrule{9-10} \cmidrule{12-13} \cmidrule{15-16} 
lutpack~\cite{Mishchenko_2008}      &  & 98.34\%      & 20.39        &                      & 83.47\%      & 64.37        &                      & 69.92\%      & 154.38        &                      & 48.95\%      & 334.79        &                      & 26.87\%      & 897.55        \\
S66~\cite{Ray_2012}      &  & 84.18\%      & 0.60        &                      & 69.24\%      & 2.57        &                      & 52.13\%      & 4.99        &                      & 37.36\%      & 6.99        &                      & 19.14\%      & 9.79        \\
66 1-SS     &  & 97.30\%      & 0.28        &                      & 82.23\%      & 1.41        &                      & 74.24\%      & 4.20        &                      & 63.06\%      & 9.39        &                      & 32.88\%      & 16.43        \\
66 M-SS   &  & 99.82\%      & 0.30        &                      & 92.94\%      & 3.08        &                      & 84.71\%      & 9.92       &                      & 63.06\%      & 9.73        &                      & 32.88\%      & 16.58        \\ \bottomrule
\end{tabularx}
\end{table*}

In this experiment, we evaluate the performance of ACD in decomposing functions by comparing it against other implementations in ABC. Specifically, we test the number of functions that can be successfully decomposed into two $6$-LUTs and the run time needed. We run this experiment on \emph{practical functions}, i.e., functions that are observable in designs and benchmarks, which include fully-, partially-, and non-DSD-decomposable functions. We extract practical functions from the EPFL benchmarks. Since the number of practical functions can be large, we classify them into $\mathcal{NPN}$-equivalence classes employing the heuristic sifting algorithm~\cite{Huang_2013, Soeken_npn}.

Table~\ref{tab:acd_success} shows the percentage of decomposable functions and the runtime for different methods and support sizes. For instance, the first column contains results for decomposing practical $7$-input functions, where $(41071)$ indicates the number of unique NPN classes collected. Each row of the table shows one ACD method. The first method \emph{lutpack}~\cite{Mishchenko_2008} performs a heuristic ACD using DSD and the Shannon's expansion, supporting up to $3$-SS variables. The second method, \emph{S66}~\cite{Ray_2012}, performs ACD using heuristic variable re-ordering supporting at most $1$-SS variable. Finally, we present two variants of our decomposition method restricted to use $2$ $6$-LUTs. One uses up to $1$-SS variable (66 $1$-SS), the other (66 M-SS) has no restrictions on the number of SS variables. The approaches described in this paper outperform the state of the art in quality for a competitive or better run time.

\subsection{Decomposition success rate for delay optimization}
\begin{table}[t]
\center
\setlength\tabcolsep{5pt}
\caption{Success ratio when decomposing practical functions into $2$ levels of $6$-LUTs with the given late-arriving variables.}\label{tab:success_rate}
\begin{tabularx}{\columnwidth}{clrrrrr}
\toprule
\textbf{N late}               & \textbf{ACD type} & \textbf{7 vars} & \textbf{8 vars} & \textbf{9 vars} & \textbf{10 vars} & \textbf{11 vars} \\ \midrule
\multirow{2}{*}{\textbf{0}} & 66 M-SS          & 99.82\%         & 92.94\%         & 84.71\%         & 63.06\%          & 32.88\%          \\
                                 & Generic           & 100.00\%        & 100.00\%        & 98.05\%         & 90.20\%          & 32.88\%          \\ \midrule
\multirow{2}{*}{\textbf{1}} & 66 M-SS          & 96.59\%         & 79.60\%         & 61.51\%         & 37.35\%          & 16.54\%          \\
                                 & Generic           & 100.00\%        & 100.00\%        & 97.57\%         & 83.23\%          & 16.54\%          \\ \midrule
\multirow{2}{*}{\textbf{2}} & 66 M-SS          & 86.22\%         & 59.78\%         & 39.28\%         & 23.74\%          & 10.95\%          \\
                                 & Generic           & 100.00\%        & 100.00\%        & 94.19\%         & 66.56\%          & 10.95\%          \\ \midrule
\multirow{2}{*}{\textbf{3}} & 66 M-SS          & 65.11\%         & 36.37\%         & 21.25\%         & 13.78\%          & 6.96\%           \\
                                 & Generic           & 93.78\%         & 86.03\%         & 76.82\%         & 44.51\%          & 6.96\%           \\ \midrule
\multirow{2}{*}{\textbf{4}} & 66 M-SS          & 36.96\%         & 17.00\%         & 8.62\%          & 7.21\%           & 4.43\%           \\
                                 & Generic           & 54.55\%         & 40.42\%         & 25.45\%         & 23.70\%          & 4.43\%           \\ \midrule
\multirow{2}{*}{\textbf{5}} & 66 M-SS          & 14.52\%         & 5.42\%          & 2.96\%          & 2.84\%           & 2.61\%           \\
                                 & Generic           & 14.52\%         & 5.42\%          & 2.96\%          & 2.84\%           & 2.61\%           \\ \bottomrule
\end{tabularx}
\end{table}
We extend the previous experiment to evaluate delay minimization using the proposed ACD method. This experiment tests the success rate of the decomposition for practical functions given delay-critical variables, which are required to be in the free set. Informally, for delay-critical variables with delay $D$, this experiment checks the existence of a decomposition with delay $D+1$. We only consider 66 M-SS and generic ACD since other known methods do not perform delay minimization using the input arrival time. For each function, we randomly generate up to $10$ unique sets of delay-critical variables and test the decomposition for each one of them. While 66 M-SS is limited to two LUTs, generic can use up to $4$ LUTs.

Table~\ref{tab:success_rate} presents the success rate based on the number of delay-critical variables, shown in the column ``N late''. The table highlights the advantage of supporting multiple BS functions. Generic ACD has a high success rate in most cases. Limitations occur when the number of delay-critical variables exceeds $3$ or the number of variables in the support is $10$ or more. Generally, the decomposition of $11$-input variables is rare. However, many $10$ input variables are still decomposable. 

\subsection{Delay-driven LUT mapping}
\begin{table*}[t]
\setlength\tabcolsep{2pt}
\centering
\caption{Comparison of delay-driven LUT mapping, LUT mapping into LUT structure ``66'', and LUT mapping using ACD.}
\footnotesize
\label{tab:lut_map}
\begin{tabularx}{\textwidth}{lXrrrrXrrrrXrrrrXrrrr}
\toprule
\textbf{Benchmark} &  & \multicolumn{4}{c}{\textbf{ABC: dch; if -K 6}} & \multicolumn{1}{l}{} & \multicolumn{4}{c}{\textbf{ABC: dch; if -s -S 66 -K 8}} & \multicolumn{1}{l}{} & \multicolumn{4}{c}{\textbf{ACD}} & \multicolumn{1}{l}{} & \multicolumn{4}{c}{\textbf{ACD; st; dch -f; if -K 6}} \\
                   &  & LUTs    & Edges    & Depth   & Time (s)   &                      & LUTs       & Edges       & Depth     & Time (s)    &                      & LUTs       & Edges      & Depth     & Time (s)    &                      & LUTs       & Edges       & Depth       & Time (s)     \\ \cmidrule{1-1} \cmidrule{3-6} \cmidrule{8-11} \cmidrule{13-16} \cmidrule{18-21} 
adder       &  & 363   & 1433   & 22    & 0.18     &  & 362      & 1465     & 20      & 0.28      &  & 383      & 1519     & 16      & 0.20     &  & 353      & 1518      & 10        & 0.39      \\
bar         &  & 1664  & 9344   & 4     & 0.44     &  & 1664     & 9344     & 4       & 0.57      &  & 1664     & 9344     & 4       & 0.47     &  & 1006     & 5274      & 4         & 0.76      \\
div         &  & 8618  & 32394  & 406   & 6.62     &  & 9107     & 33665    & 397     & 13.42     &  & 11644    & 44496    & 326     & 7.16     &  & 9068     & 39167     & 271       & 21.19     \\
hyp         &  & 58393 & 239097 & 1864  & 5.43     &  & 61701    & 247699   & 1840    & 31.82     &  & 65615    & 264998   & 1396    & 11.13    &  & 61769    & 263254    & 1034      & 19.76     \\
log2        &  & 9712  & 43562  & 58    & 17.05    &  & 10172    & 44943    & 58      & 30.06     &  & 10313    & 46365    & 56      & 17.81    &  & 9429     & 42533     & 57        & 39.09     \\
max         &  & 831   & 3804   & 14    & 0.37     &  & 840      & 3668     & 14      & 0.63      &  & 1211     & 5578     & 12      & 0.42     &  & 871      & 4277      & 11        & 1.39      \\
multiplier  &  & 7383  & 34137  & 36    & 6.01     &  & 7334     & 32781    & 36      & 12.11     &  & 7693     & 35798    & 33      & 6.82     &  & 6800     & 31705     & 31        & 13.32     \\
sin         &  & 1928  & 8445   & 30    & 1.31     &  & 1948     & 8463     & 30      & 4.94      &  & 2052     & 8913     & 29      & 1.50     &  & 1830     & 8178      & 30        & 2.91      \\
sqrt        &  & 7515  & 29573  & 663   & 4.17     &  & 7972     & 30610    & 638     & 12.66     &  & 10156    & 38558    & 519     & 4.73     &  & 9292     & 36030     & 476       & 8.77      \\
square      &  & 4122  & 17319  & 23    & 1.98     &  & 4165     & 17547    & 22      & 3.91      &  & 4107     & 17924    & 18      & 2.22     &  & 4118     & 18285     & 14        & 5.15      \\
arbiter     &  & 1833  & 8982   & 6     & 1.64     &  & 1879     & 8836     & 6       & 2.02      &  & 1850     & 8987     & 6       & 1.70     &  & 2037     & 8780      & 6         & 3.33      \\
cavlc       &  & 137   & 707    & 4     & 0.13     &  & 104      & 491      & 4       & 0.56      &  & 137      & 707      & 4       & 0.15     &  & 123      & 655       & 4         & 0.20      \\
ctrl        &  & 30    & 133    & 2     & 0.07     &  & 28       & 127      & 2       & 0.08      &  & 30       & 133      & 2       & 0.08     &  & 29       & 126       & 2         & 0.08      \\
dec         &  & 287   & 684    & 2     & 0.09     &  & 287      & 1404     & 2       & 0.1       &  & 287      & 684      & 2       & 0.10     &  & 284      & 816       & 2         & 0.12      \\
i2c         &  & 312   & 1360   & 3     & 0.16     &  & 306      & 1316     & 3       & 0.36      &  & 319      & 1378     & 3       & 0.19     &  & 297      & 1329      & 3         & 0.27      \\
int2float   &  & 52    & 258    & 3     & 0.08     &  & 46       & 205      & 3       & 0.18      &  & 52       & 258      & 3       & 0.09     &  & 50       & 251       & 3         & 0.11      \\
mem\_ctrl   &  & 11037 & 48812  & 18    & 10.24    &  & 10830    & 46368    & 18      & 31.67     &  & 11232    & 49483    & 17      & 11.40    &  & 10398    & 45793     & 16        & 20.57     \\
priority    &  & 178   & 725    & 6     & 0.11     &  & 182      & 736      & 6       & 0.18      &  & 185      & 736      & 6       & 0.12     &  & 171      & 698       & 6         & 0.17      \\
router      &  & 89    & 285    & 4     & 0.09     &  & 61       & 283      & 4       & 0.14      &  & 92       & 290      & 4       & 0.09     &  & 89       & 279       & 4         & 0.12      \\
voter       &  & 1838  & 8596   & 13    & 2.23     &  & 1784     & 8624     & 13      & 4.14      &  & 1838     & 8583     & 13      & 2.32     &  & 1777     & 8426      & 13        & 4.82      \\ \midrule
Improvement &  &       &        &       &          &  & 2.57\%   & -2.57\%  & 1.04\%  &           &  & -8.13\%  & -7.87\%  & 7.52\%  &          &  & 2.20\%   & -0.30\%   & 12.39\%   &           \\
Total       &  &       &        &       & 58.40    &  &          &          &         & 149.83    &  &          &          &         & 68.70    &  &          &           &           & 142.52       \\ \bottomrule
\end{tabularx}
\end{table*}

Table~\ref{tab:lut_map} compares four technology mapping strategies for delay minimization during mapping into $6$-LUTs, assuming a unit-delay model. Each strategy takes the baseline as an input and computes structural choices before mapping. Structural choices have not been used for the benchmark \emph{hyp} due to a known bug in ABC. The proposed method is compared against standard LUT mapping and mapping into LUT structures. Command \emph{ACD} denotes our mapper with Boolean decomposition using the sequence ``\texttt{dch; if -Z 6 -K 8}''. We do not compare against~\cite{Ray_2012} and~\cite{Mishchenko_2008} because those methods do not support delay minimization. Furthermore, we do not compare against the recent mapper with gate decomposition based on bin-backing~\cite{Fan_2023}. Nevertheless, the mapper in~\cite{Fan_2023} would improve the average delay of ABC \texttt{if} by only $0.31$\%.

Mapping into LUT structure ``$66$'' composed of two 6-LUTs, which is a SAT-based version of structural ACD, reduces depth by $1.04$\% and the area by $2.57$\% on average, at the cost of increasing the number of edges by $2.57$\%. The proposed LUT mapping with ACD improves the depth of the LUT network by $7.52$\% on average while increasing the number of LUTs and edges by $8.13$\% and $7.87$\%, respectively. 

Note that most of the improvement is concentrated in the first $10$ benchmarks since others are already close to their best known depth~\cite{EPFL_best}. For $4$ of them, the delay reduction exceeds $20$\% and is up to $27.27$\%. Practically, part of the area increase can be reduced by area-recovery methods~\cite{Mishchenko_2008, Mishchenko_2011, Schmitt_2018}, using delay relaxation, or by an additional mapping step applied after ACD. The rightmost strategy performs the latter option. The LUT count and edge count are reduced considerably, leading to an area improvement of $2.20$\%, compared to traditional technology mapping with choices. Also, the logical depth further decreases up to $54.55$\%. Specifically, the result after ACD is used as a choice to improve the next round of technology mapping because choices extracted from mapping with ACD are more structurally suited to delay-oriented mapping, compared to the original AIG. Moreover, structural choices help reduce the area over the non-critical paths. Note that a second mapping round does not provide practical benefits if applied to the default LUT mapper (leftmost column) since the network after deriving the AIG is structurally similar to the baseline. Furthermore, benchmark \emph{hyp} is noticeably improved by remapping both in area and delay without using structural choices. Regarding the run time, mapping with ACD is faster than mapping into LUT structures while being more general. 

\subsection{EPFL synthesis competition}
\begin{table}[t]
\setlength\tabcolsep{2pt}
\centering
\caption{LUT mapping in the EPFL synthesis competition.}
\label{tab:best}

\begin{tabularx}{\columnwidth}{lXrrXrrXrr}
\toprule
\textbf{Benchmark} & \textbf{} & \multicolumn{2}{c}{\textbf{Best~\cite{EPFL_best}}} &  & \multicolumn{2}{c}{\textbf{dch -f; if -K 6}}                                                        & \textbf{} & \multicolumn{2}{c}{\textbf{dch -f; if -Z 6 -K 10}}                                                       \\
                                    &           & LUTs            & Depth          &  & LUTs                                              & Depth                                           &           & LUTs                                                 & Depth                                             \\ \cmidrule{1-1} \cmidrule{3-4} \cmidrule{6-7} \cmidrule{9-10} 
adder                               &           & 347             & 5              &  & 360                                               & 6                                               &           & 445                                                  & {\color[HTML]{3531FF} 5}   \\
bar                                 &           & 512             & 4              &  & {\color[HTML]{3531FF} 512} & {\color[HTML]{3531FF} 4} &           & {\color[HTML]{3531FF} 512}    & {\color[HTML]{3531FF} 4}   \\
div                                 &           & 25318           & 175            &  & 23461                                             & 192                                             &           & 31526                                                & {\color[HTML]{3531FF} 175} \\
hyp                                 &           & 182723          & 483            &  & 122394                                            & 511                                             &           & {\color[HTML]{009901} 154903} & {\color[HTML]{009901} 473} \\
log2                                &           & 8617            & 52             &  & 8778                                              & 60                                              &           & 9613                                                 & {\color[HTML]{009901} 51}  \\
max                                 &           & 1114            & 6              &  & 1113                                              & 7                                               &           & 1250                                                 & {\color[HTML]{3531FF} 6}   \\
multiplier                          &           & 7785            & 25             &  & 6839                                              & 28                                              &           & {\color[HTML]{009901} 6903}   & {\color[HTML]{3531FF} 25}  \\
sin                                 &           & 680530          & 10             &  & 1820                                              & 33                                              &           & 2379                                                 & 27                                                \\
sqrt                                &           & 29593           & 162            &  & 30945                                             & 172                                             &           & 41626                                                & {\color[HTML]{009901} 156} \\
square                              &           & 3732            & 10             &  & 4189                                              & 11                                              &           & 4275                                                 & {\color[HTML]{3531FF} 10}  \\ \bottomrule
\end{tabularx}
\end{table}
This experiment shows that mapping using ACD can improve well-optimized LUT networks, resulting in best known results for $4$ benchmarks in the EPFL synthesis competition. The previous best results were obtained using a portfolio of heavy logic optimizations applied to various representations, such as AIGs and LUT networks. In recent years, results have been further improved using \emph{design-space exploration}~(DSE) techniques that incrementally generate optimization scripts.


We obtain the optimized AIGs by repeatedly running the script used in the baseline of Table~\ref{tab:lut_map} along with additional delay-oriented AIG commands in ABC. From the obtained AIG, we compare traditional LUT mapping with choices to LUT mapping with ACD. Notably, results from the traditional mapper are quite far from the best results. This observation shows, as expected, that our technology-independent optimization finds worse AIGs than those used to obtain the best results. However, LUT mapping with ACD matches or improves the depth for almost all benchmarks. The improved benchmarks are \emph{hyp}, \emph{log2}, \emph{multiplier}, and \emph{square}. Remarkably, our method reduces the depth of \emph{hyp} by $10$ levels, compared to the state of the art, while reducing area by $15$\%. In the benchmark \emph{multiplier}, our result matches the depth but improves the number of LUTs. Benchmark \emph{sin} is the only one where there is a large gap compared to the best result. In particular, the best result for \emph{sin} requires significant logic duplication that is not performed in our synthesis flow.
Contrarily to many other methods used to produce the best results, our results in Table~\ref{tab:lut_map} are obtained directly by LUT mapping without employing post-mapping optimization. 


\section{Conclusion} \label{sec:conclusion}
This work proposes a novel formulation of Ashenhurst-Curtis decomposition~(ACD) that enables efficient technology mapping and post-mapping resynthesis. The algorithm is truth-table-based and works for any size of the free set, bound set, and shared set, which makes it well-suited for delay optimization. We have shown that the proposed Boolean decomposition improves state-of-the-art in the decomposition quality with a competitive runtime. We have implemented and integrated the proposed method into a delay-driven LUT mapper. The experiments have shown that LUT mapping with ACD can improve the average delay by $12.39$\%, compared to the traditional structural LUT mapping with choices. Furthermore, the proposed approach has produced best results for $4$ test cases in the EPFL synthesis competition.




\section*{Acknowledgments} \label{sec:akw}
This research was supported by the SNF grant ``Supercool: Design methods and tools for superconducting electronics'', 200021\_1920981, and Synopsys Inc.

\bibliographystyle{IEEEtran}
\bibliography{bibliography}




\end{document}